\def\cE{{\cal E}}
\def\cH{{\cal H}}

\def\Tr{\hbox{Tr}\,}
\def\ket#1{\left|\, #1\right\rangle}
\def\bra#1{\left\langle #1\right|}

\documentclass{article}
\usepackage{hyperref}
\title{Nonlinear Quantum Gravity}
\author{George Svetlichny\footnote{Departamento de Matem\'atica, Pontif\'{\i}cia Universidade Cat\'olica, Rio de Janeiro, Brazil \newline
svetlich@mat.puc-rio.br \hfill \url{http://www.mat.puc-rio.br/\~svetlich}}}
\begin{document}
\maketitle

\begin{abstract}
Nonlinear quantum mechanics at the Planck scale can produce nonlocal effects contributing to resolution of singularities, to cosmic acceleration, and modified black-hole dynamics, while avoiding the usual causality issues. 
\end{abstract}

\section{Introduction} We explore here some possible consequences for quantum gravity if quantum mechanics becomes nonlinear at the Planck scale, and speculate that black hole dynamics may be linked to the accelerated expansion of the universe through a connection between short and long wavelength modes. Space limitations only allow for limited discussion and  references on nonlinear quantum mechanics for which  see\cite{svenlqmp,sverl}. Now the principal characteristic of nonlinearity can be summarized as follows: {\sl generically, entangled systems become causal channels.} To appreciate this fact consider a composite system described by a tensor product Hilbert space \(\cH_1\otimes \cH_2\) and a Schr\"odinger time evolution given by \(i\partial_t\Psi=H\Psi\) where \(H\) is a not necessarily linear operator and \(\Psi\in \cH_1\otimes \cH_2\). Among such evolutions there are those known as {\em  separating\/} which means that product vectors evolve as product vectors, that is if \(\Psi(0)=\Psi_1(0)\otimes \Psi_2(0)\) then  \(\Psi(t)=\Psi_1(t)\otimes \Psi_2(t)\) and each \(\Psi_i(t)\) evolves by its own Schr\"odinger time evolution \(i\partial_t\Psi_j=H_j\Psi_j\), \(j=1,2\). Separable systems for distinguishable parts have been fully classified by Goldin and Svetlichny \cite{GGGS:separation}.  One has then \(H=H_1+H_2+K\) where \(K\) is an operator that vanishes on product states. Separability is a nonlinear expression of lack of interaction. Now if \(H\) is linear and separable, then \(K=0\) and even if \(\Psi\) is not a product state, its partial trace in \(\cH_1\), \(\rho_1=\Tr_2{\ket \Psi\bra\Psi}\) satisfies the von Neumann evolution equation
 \(i\partial_t\rho_1=[H_1,\rho_1]\),
 and so all initial states \(\Psi(0)\) that have the same partial trace \(\rho_1(0)\) lead to the same local evolution, independently of what the operator \(H_2\) is and of the further details of entanglement of the two parts represented by \(\Psi(0)\). The same statement of course  holds for the other partial trace \(\rho_2\). If now the \(H_i\) are nonlinear, then even if \(K=0\), the partial traces generically do not have independent evolution. This means that entanglement in \(\Psi(0)\) lead to a causal connection between the  two parts even though the evolution is ostensibly noninteractive. This has been used as an argument against quantum nonlinearity since now one can show that EPR-type correlations along with the usual hypothesis of state collapse due to measurements can be used to send superluminal signals, calling into question relativistic causality. Since the idea of measurements being performed during the Planck epoch of the evolution of the universe is somewhat bizarre, the consequence of this situation for quantum gravity has not been properly appreciated. Decoherence however shares many properties of measurement and so if we consider \(
 \cH_i=\cH_{S_i}\otimes \cH_{\cE_i}\) where \(\cH_{S_i}\) are Hilbert spaces of some quantum systems inside an environment describe by Hilbert spaces \(\cH_{\cE_i}\) and in which decoherence occurs through some nonlinear quantum process, then if \(\Psi(0)\) is entangled, generically the decoherence process in one part will causally influence that in the other part, again even though the overall nonlinear Hamiltonian is separating. One cannot deny the importance of decoherence in the early evolution of the universe, and so this type of causal channel would be quite relevant. There have been many attempts to circumvent the causality issue by a deeper analysis of the measurement process and its relation to evolution, introducing appropriate modifications or reinterpretations of both. On the one hand it does seem ironic that by this one is attempting to  eliminate precisely the main distinguishing characteristic of nonlinearity; on the other hand the ubiquity of entangled systems means that such causal channels must proliferate in wide variety of circumstances (for example,  entanglement of spin and orbital angular momentum states in atomic physics) which  may or may not be causally problematic. Ironically, again, there seems to be no considerations of these other channels in the literature. With nonlinear quantum mechanics causal channels abound beyond anything conceived in linear theory. This situation turns nonlinear quantum systems radical, introducing effects which to many are unwelcome and often leads to a rejection, practically off-hand, of nonlinear theories.

 There are however cogent reasons for considering nonlinear quantum mechanics. Surprisingly enough, non-linear quantum mechanics appears through {\em  linear\/} representations of the diffeomorphism group.\cite{doegol}
 As a consequence one can expect nonlinear quantum processes to unexpectedly show up in any theory that says its both quantum and geometric. The lesson is: {\sl nonlinear quantum mechanics is intrinsically associated with quantum geometry and ignoring it may not be wise.\/} Another reason is more conceptual. Take the usual dictum of general relativity: {\em  Space-time tells matter how to move; matter tells space-time how to curve\/}, and perform a {\em  verbal\/} quantization of general relativity introducing the adjective ``quantum" for space-time and matter: {\em  Quantum space-time tells matter how to move; quantum matter tells space-time how to curve\/}. \footnote{Maybe ``move" and ``curve" are no longer appropriate words, but we can remove this ``anomaly" with some words such as ``behave" and ``be". } This seeming natural, {\em  relational\/}, viewpoint leads to a nonlinear quantum theory as there is a back reaction of matter on its own dynamics. The prevailing {\em absolutist\/} position is that {\em  linear\/} quantum mechanics tells both space-time and matter how to be. Of course only experiment can decide who's right.

\section{Nonlinear Quantum Effects}

So what about problems like causality? Entanglement causality channels abound, eliminate effects in some, others are still there. Our main concern here    is not whether causal channels exist or not (they are ubiquitous) but determining which ones are strong and which are weak, that is which modify substantially linear behavior, and which modify it below practical thresholds. Experiments suggest that at low energies nonlinear effects are \(\le 10^{-20} \) times smaller than linear effects. Pushing this further we can speculate that if effects only appear at Planck energies they may not cause problems as then space-time itself becomes quantum with ill-defined causal structure making it nonsensical to talk about its violation. The next question is: can effects be large at the Plank scale and still be suppressed at low energies?

A hint toward an answer comes from considering the Doebner-Goldin (DG) equation, which is the nonlinear evolution connected to representations of the diffeomorphism group. Explicitly the one-particle equation is:
\[i\hbar\partial_t\psi_s = F_s\psi_s = -\frac{\hbar^2}{2m_s}\nabla^2 \psi_s +
iD_s\hbar\left(\nabla^2\psi_s +
\frac{|\nabla\psi_s|^2}{|\psi_s|^2}\psi_s\right)+R_s(\psi)\psi,\]
where \(s\) labels the particle's species, \(D_s\) is a physical constant and \(R_s(\psi)\) is {\em  real\/} and
complex homogeneous of degree zero: \(R_s(z\psi)=R_s(\psi)\).

Using a zero-momentum two-particle (\(a\) and \(b\)) EPR state \(\phi\), one finds that the difference, to first order  in \(t\), of the matrix element \((\psi(t),B\psi(t))=t\Delta_1(B|p,q)+O(t^2)\) of an observable \(B\) on particle \(b\) between post-position (\(q\)) and post-momentum (\(p\)) measurement (at \(t=0\)) upon particle  \(a\)  is, asymptotically (as \(s\to\infty\)):
\[
\Delta_1(B|p,q) = 4sn D_b(\phi,B\phi) + O(1),
\]
where we use gaussian position states  \[\delta^{(s)}(y) = ({s \over \pi})^{n/2}e^{-s y^2}.\]

One sees then that {\em  even if  \(D_b\) is extremely small, under extreme localization the effect can be large\/}.

The effect is relatively larger for longer wavelength modes. Roughly
\[\frac{\hbox{\small NONLINER EFFECT}}{\hbox{\small LINEAR EFFECT}}\approx \frac{D_b}{\hbar/2m} sL^2\]  where  \(L\) is the wavelength of the affected mode.

Thus the DG equation suggests a nonlinear quantum gravity effect coupling short (Planck) wavelengths to long (possibly Hubble) wavelengths.
\cite{sveamp}

One sees then that if Planck and Hubble modes are entangled, nonlinear effects can be large and still be suppressed at normal energies.

Is there any hope though of seeing a signature of nonlinear quantum mechanics? If nonlinear effects are  suppressed at low energies and only at Planck energies are comparable to linear effects, it does seem discouraging. However it has been recently recognized that some form of ``quantum gravity phenomenology" (cosmic rays propagation, tests of  violations of lorentz invariance, CTP, or unitarity) is possible. Such effects are proposed within linear quantum mechanics, but can also, alternatively, be construed as tests of nonlinear quantum theories which through other mechanisms lead to the same sort of effects. Unfortunately these tests do not distinguish between the two types of theories.
Just as elliptical planetary orbits can  be explained by enough epicycles, effects arising from nonlinear quantum mechanics can most likely be explained by more elaborate linear theories. Because of its radical nature, a nonlinear theory would only be accepted if it bring greater simplicity (such as ellipses vs. epicycles) or there is some observation for which a linear quantum theory explanation is not readily forthcoming. Such may be the case in cosmology.

\section{Nonlinear Quantum Gravity}
One possible nonlinear effect has to do with the accelerated expansion of the universe and the problem of space-time singularities, which are generically present in general relativity  under some widely accepted hypotheses. The singularity theorems depend on the satisfaction of so-called energy conditions on the stress-energy tensor \(T_{\mu\nu}\). Two such are the {\em  weak\/} energy condition \(T_{\mu\nu}U^\mu U^\nu\ge 0\) for \(U\) time-like,
and the {\em  dominant\/} energy condition which requires in addition that
\(T^\mu{}_\nu U^\nu\) be not space-like.

Since quantum gravity should resolve the problem of singularities, a
semiclassical theory of quantum gravity should violate some energy conditions.
Now it's been known for some time that quantum field theory violates energy conditions, but such violations are limited by so-called quantum inequalities and so far this type of violation has not been shown to avoid the formation of singularities. On the empirical side, the
accelerated expansion of the universe possibly does violate the dominant energy condition as the observational data concerning the so-called ``dark energy" component of the universe is consistent with an equation of state
\(p=w\rho\) with \(w<-1\),
where \(\rho>0\) is energy density and \(p<0\) is pressure. Such a dark energy has been dubbed ``phantom energy" and has a series of remarkable properties, one of which is that black holes accreting phantom energy can lose mass instead of gaining it.\cite{babi.etal089}
Ordinary linear quantum field theory can even violate the dominant energy condition on a {\em  cosmic\/} scale, but again such violations are limited\cite{onewoo} and probably cannot readily account for the present acceleration.

The space-time region outside a black hole event horizon is about as causally remote from the region inside as can be imagined. The event horizon is maintained by a metric that has a central singularity. According to present ideas, as one approaches the singularity, one enters a Planck regime and so only some form of quantum gravity can give account of the physics, in particular avoiding a true singularity. Near such a would-be singularity a semiclassical theory will violate some of the usual energy conditions. Since phantom energy violates the dominant energy condition and since phantom energy can also by accretion diminish the central singularity of a black hole through a reduction of its mass, one can speculate that quantum gravity uses phantom energy to resolve singularities. Phantom energy in our universe seems to be tied to Hubble-scale processes, but singularities such as in the centers of black holes are Planck-scale. From what was said above it seems possible that nonlinear quantum mechanics can by relating these two scales be responsible for the presence of phantom energy and at the same time resolve (at least some) space-time singularities.
One comes therefore to the strange hypothesis that black holes may be responsible for the presence of phantom energy. Nonlinear quantum mechanics, which can relate short- and long-wavelengths processes, would entangle them, this entanglement becomes a causal channel between the interior and exterior regions to the event horizon, the  would-be singularity would act as a position measuring device (read ``decoherence environment") for short-wavelength and  project the long-wavelength partner modes into Hubble-size phantom energy states which by accretion diminish the black-hole mass and in time remove the would-be singularity. This scenario is explained in greater detail in \cite{svebhph}. Space limitation here allows us only to address a few  relevant issues.

There is one cosmological observational consequence of such a hypothesis. The universe today is dominated by dark energy and dark matter. Assume the energy is phantom, and assume that the dark energy is produced by nonlinear quantum processes within black holes as they accrete matter or not, then at some time in the cosmic epoch there would be energy flow from the matter sector to the dark energy sector. Such a two-sector situation would be modelled by the following  FRW equations  valid during a certain time period of cosmic evolution, including the current one:

\begin{eqnarray*} \label{fried}
\left(\frac{\dot a}a\right)^2 = \frac{\kappa_0^2}3(\rho_m+\rho_{\rm ph}),\\ \label{darkcont}
\dot\rho_m +3H\rho_m = - b\rho_m,\\ \label{phantomcont}
\dot\rho_{\rm ph} +3\gamma H\rho_{\rm ph}=  b\rho_m.
\end{eqnarray*}
here \(\rho_m\) and \(\rho_{\rm ph}\) are the matter and phantom energy densities, \(\kappa_0=8\pi G\), \(H=\dot a/a\) is the Hubble parameter,  \(\gamma=w+1<0\), and \(b\) represents the coupling of matter to phantom energy mediated by black-hole singularities.

The above equations fall under the broad category of interacting dark energy models. The prevailing hypothesis seems to be that energy flows from the dark energy sector to the dark matter one. Our hypothesis that at some epoch the flow is in the other directions is a distinguishing feature. Introducing \(\rho=\rho_m+\rho_{\rm ph}\) and \(\Omega_m=\rho_m/\rho\) one can deduce

\begin{equation}\label{b}
 b=  3wH(1-\Omega_m)-\frac{\dot \Omega_m}{\Omega_m},
\end{equation}
the right hand side of of which can in principle be evaluated by empirical data.

The only definite prediction we can make about the \(b\) function is that it must be positive at some time. It is positive if energy flows from matter to phantom, and negative if in the other direction. If black-hole mediation of phantom energy production is true, the sign of the right hand side of (\ref{b}) should be positive during some part of the universe's evolution after the radiation dominated era. Whether it should still be positive at the present coincidental moment when dark matter and dark energy have densities of the same order of magnitude is not a-priori clear as phantom energy could also be transferring energy to dark matter by some mechanism making the net energy flow from phantom to matter.
 Presently considered interacting dark energy theories apparently all consider negative \(b\), hence positivity could very well be a true signature of nonlinear quantum mechanics which can naturally accommodate it. A failure will impose non-trivial constraints on any nonlinear version of quantum gravity  eliminating some of the more striking aspects the theory might have otherwise.
In any case, the right-hand side of (\ref{b}) is an important empirical datum in our search for a better understanding of quantum features of space-time.

Though there is data concerning the evolution of dark matter \cite{bacoetal}, it doesn't seem to be sufficiently precise to determine the derivative \(\dot \Omega_m\) with any accuracy. Szyd{\l}owski \cite{szyd} argues for a negative sign for \(b_0\) (the value now) based on SNIa supernovae observations, but again, given the uncertainties in the data, a positive sign cannot be entirely ruled out.

A vision of quantum gravity that would allow for the above type of effect can be formulated in analogy with thermodynamics.  Think of space-time as a ferromagnet. The metric would correspond to magnetization, Planck energy to critical energy (temperature),  and the region near a singularity as  the disordered phase. This phase would not be metrically related to the ordered phase (since order {\em  is\/} metricity) where a metric structure exists, allowing thus for non-local effects to be mediated by would-be classical singularities. A ferromagnetic-type model for regions near a black hole singularity has also been proposed by Ashtekar and Bojowald.\cite{as-bo} In our view, nonlinear quantum processes would be relevant near Planck energy (\(\approx 10^{19}\,\hbox{Gev}\)), and because of short-long wavelength entanglement also at Hubble energy (\(\approx 10^{-35}\,\hbox{ev}\)). Between these energies ordinary linear quantum mechanics holds sway, and beyond in either direction is the disordered phase, completing thus a full circle.

We can finally address the causality issue of nonlinear quantum mechanics. A causal link through entangled Planck and Hubble wavelengths does not create the usual causality problems since to detect a mode of wavelength \(L\) requires an apparatus  acting on a time scale of order \(L/c\). Any Hubble size mode created non-locally by a Planck-scale process could only be detected at a time that is already future time-like to the creation event, avoiding, in a new way, the usual causality violation problem claimed of nonlinear quantum mechanics. Nonlinear quantum gravity could thus escape the problem of an abundance of causal channels due to entanglement, limiting them only to the two far ends of the energy spectrum (Planck and Hubble) while at the same time resolve in a novel way the singularity issues of classical general relativity.

\section{Things to Do}

The lack of a clear mathematical theory and empirical data makes the above exposition highly speculative. Certain avenues of future research naturally present themselves:

\begin{enumerate}
\item  Improve the cosmological model and compare to observations, especially concerning the direction of energy flow between  matter and dark energy. The direction of this flow during the epoch when these two sectors dominate the dynamics of the universe is an important datum for nonlinear quantum theory.
\item Develop the decoherence theory for nonlinear evolution, especially for the DG equation. This would allow the causality questions that have so far been limited to measurement situations to be adequately addressed  in contexts where measurement activity does not make sense.
\item Investigate to what extent nonlinear quantum mechanics violates  quantum inequalities obeyed by linear quantum mechanics. Violation of the energy conditions by ordinary quantum field theory is limited by quantum inequalities to such an extent that space-time singularities (such as wormhole collapse) are not avoided. A nonlinear theory could be more effective in this regard.
\item Investigate further the apparent connection of nonlinear quantum mechanics and noncommutative spaces as suggested by T.~P.~Singh, {\sl et. al.\/}\cite{sinetal} and T.~P.~Singh.\cite{singh} This would further our understanding of the connection between nonlinear quantum mechanics and geometry, already implied by diffeomorphism group representations, and give further support to the idea that nonlinear quantum mechanics may have something to do with our world.
\end{enumerate}
\section*{Acknowledgements}
This research was partially supported by the Conselho Nacional de Desenvolvimento Cien\'{\i}fico e Tecnol\'ogico (CNPq).


\begin{thebibliography}{xxx}
\bibitem{svenlqmp}George Svetlichny, ``Nonlinear Quantum Mechanics at the Planck Scale",
quant-ph/0410230 (Talk given at the International Quantum Structures 2004 meeting, to appear  in th {\sl International Journal of Theoretical Physics})
\bibitem{sverl}George Svetlichny, ``Informal Resource Letter - Nonlinear quantum mechanics on arXiv up to August 2004", quant-ph/0410036.
\bibitem{GGGS:separation} G.~A.~Goldin and  G.~Svetlichny,  {\sl J. Math. Phys.} \textbf{35}, 3322 (1994).
\bibitem{doegol}H.-D.~Doebner and G.~A.~Goldin {\sl Phys.\ Lett.\/}\ A {\bf 162} 397 (1992).

\bibitem{sveamp} G.~Svetlichny, ``Amplification of Nonlocal Effects in Nonlinear Quantum Mechanics Under Localization",
quant-ph/0410186.
\bibitem{babi.etal089} E.~Babichev, V.~Dokuchaev and Yu.~Eroshenko,  {\sl Physical Review Letters\/} \textbf{93}, 021102 (2004), gr-qc/0402089.

\bibitem{onewoo} V.~K.~Onemli and R.~P.~Woodard, ``Quantum effects can render w<-1 on cosmological scales", {\sl Physical Review \/} \textbf{D70}, 107301 (2004), gr-qc/0406098.

\bibitem{svebhph}George Svetlichny, ``Do Black Holes Produce Phantom Energy?", astro-ph/0503325.


\bibitem{bacoetal}Bacon D J  {\sl et. al.\/} 2004 Evolution of the Dark Matter Distribution with 3-D Weak Lensing {\it Preprint\/} astro-ph/0403384

\bibitem{szyd}Szyd{\l}owski M 2005 Cosmological Model with Energy Transfer {\it Preprint\/} astro-ph/0502034.

\bibitem{as-bo}  A.~Ashtekar, M.~Bojowald,
{\sl Classical and Quantum Gravity\/}  \textbf{22}, 3349 (2005).

\bibitem{sinetal}T.P.Singh, Sashideep Gutti, Rakesh Tibrewala, ``Quantum Mechanics without spacetime V - Why a quantum theory of gravity should be non-linear-", gr-qc/0503116.

\bibitem{singh}T.~P.~Singh ``Quantum mechanics without spacetime: a case for noncommutative geometry", gr-qc/0510042.
\end{thebibliography}
\end{document}